\documentclass[runningheads]{llncs}

\usepackage{lipsum}
\usepackage{graphicx}
\usepackage{hyperref}
\usepackage{booktabs}
\newcommand{\ra}[1]{\renewcommand{\arraystretch}{#1}}

\usepackage[tableposition=top]{caption}

\begin{document}
%
%
\title{Evaluating U-net Brain Extraction for Multi-site and Longitudinal Preclinical Stroke Imaging}
\titlerunning{Evaluating U-net Brain Extraction for ...}
%

\author{Erendiz Tarakci$^1$
\and Joseph Mandeville$^2$
\and Fahmeed Hyder$^{3,4}$ 
\and Basavaraju G. Sanganahalli$^4$
\and Daniel R. Thedens$^5$
\and Ali Arbab$^6$
\and Shuning Huang$^7$
\and Adnan Bibic$^8$
\and Jelena Mihailovic$^3$
\and Andreia Morais$^9$
\and Jessica Lamb$^{11}$
\and Karisma Nagarkatti$^{11}$
\and Marcio A. Dinitz$^{12}$
\and Andre Rogatko$^{12}$
\and Arthur W. Toga$^1$
\and Patrick Lyden$^{10,11}$
\and Cenk Ayata$^9$
Ryan P. Cabeen$^{*,1}$}
\authorrunning{Cabeen et al.}

\institute{
$^1$Laboratory of Neuro Imaging, USC Mark and Mary Stevens Imaging and Informatics Institute, Keck School of Medicine of USC; Los Angeles, CA USA,
$^2$Athinoula A. Martinos Center for Biomedical Imaging, Department of Radiology, Massachusetts General Hospital, and Harvard Medical School, Charlestown, Massachusetts, USA,
$^3$Departments of Biomedical Engineering,
$^4$Departments of Radiology and Biomedical Imaging, Yale University, New Haven, CT USA,
$^5$Carver College of Medicine, and Department of Epidemiology, University of Iowa,
$^6$Medical College of Georgia, Augusta University, Augusta, GA, USA,
$^7$Department of Diagnostic and Interventional Imaging, McGovern Medical School, University of Texas Health Science Center at Houston, Houston, TX,
$^8$Department of Anesthesiology and Critical Care Medicine, Johns Hopkins University; Baltimore, MD USA,
$^9$Department of Radiology, Department of Neurology, Harvard Medical School, Massachusetts General Hospital, Charlestown, MA, United States
$^{10}$Department of Neurology,
$^{11}$Department of Physiology and Neuroscience, Zilkha Neurogenetic Institute,, Keck School of Medicine at USC; Los Angeles, CA USA
$^{12}$Biostatistics and Bioinformatics Research Center, Samuel Oschin Comprehensive Cancer Center, Cedars-Sinai Medical Center, Los Angeles, CA, United States,\\
$^{*}$Corresponding author: \url{rcabeen@loni.usc.edu}}

\maketitle              

\begin{abstract}
Rodent stroke models are important for evaluating treatments and understanding the pathophysiology and behavioral changes of brain ischemia, and magnetic resonance imaging (MRI) is a valuable tool for measuring outcome in preclinical studies.  Brain extraction is an essential first step in most neuroimaging pipelines; however, it can be challenging in the presence of severe pathology and when dataset quality is highly variable.  Convolutional neural networks (CNNs) can improve accuracy and reduce operator time, facilitating high throughput preclinical studies. As part of an ongoing preclinical stroke imaging study, we developed a deep-learning mouse brain extraction tool by using a U-net CNN.  While previous studies have evaluated U-net architectures, we sought to evaluate their practical performance across data types.  We ask how performance is affected with data across: six imaging centers, two time points after experimental stroke, and across four MRI contrasts.   We trained, validated, and tested a typical U-net model on 240 multimodal MRI datasets including quantitative multi-echo T2 and apparent diffusivity coefficient (ADC) maps, and performed qualitative evaluation with a large preclinical stroke database (N=1,368). We describe the design and development of this system, and report our findings linking data characteristics to segmentation performance.  We consistently found high accuracy and ability of the U-net architecture to generalize performance in a range of 95-97\% accuracy, with only modest reductions in performance based on lower fidelity imaging hardware and brain pathology.  This work can help inform the design of future preclinical rodent imaging studies and improve their scalability and reliability.


\keywords{stroke \and preclinical MRI \and segmentation \and deep learning \and artificial intelligence \and u-net architecture \and rodents \and evaluation}
\end{abstract}
\section{Introduction}

Rodent models are widely used in preclinical studies to evaluate treatments and understand pathophysiology and behavioral changes associated with a variety of brain diseases and injuries, including ischemia \cite{Lo2003-nn} and traumatic injury \cite{cabeen2020computational}. Magnetic Resonance Imaging (MRI) is an emerging, valuable tool in establishing tissue readout measures in such preclinical studies \cite{chamorro2021future}. Compared to histopathology and microscopy, MRI has the advantage of being more readily standardized, acquired at multiple time points, applied to measure multiple tissue characteristics simultaneously, and it has a more direct translation path to treating patients.  Preclinical MRI analysis pipelines often involve a complex sequence of processes, but an essential initial step of most pipelines is brain extraction, which isolates the brain from the skull and head. A successful brain extraction tool should be able to perform on a large data set that results from a variety acquisition setups and image contrasts. A variety of robust brain extraction tools have been developed for human MRI \cite{ashburner2012spm} \cite{smith2000bet}; however, these tools are not directly applicable to rodent data due to structural differences in the tissue and differences in image acquisition, e.g. field strength and imaging sequences.  For rodent models, there is no standardized automated brain extraction tool that performs as well as those built for human models, perhaps due to heterogeneity of rodent imaging approaches and a relatively smaller demand.

Traditional brain extraction tools for rodent MRI have demonstrated successful automation with high accuracy \cite{oguz2014rats} \cite{chou2011robust} \cite{liu2020automatic} \cite{ruan2021automated}; however, there remain practical limitations due a need for parameter optimization based on specific sequences used and a high failure rate when applying tools on new datasets, particularly due to pathology in preclinical trials involving brain trauma.  State-of-the-art approaches for brain extraction in non-human models now use convolutional neural networks (CNNs) \cite{hsu2020automatic} \cite{de2021automated} \cite{pontes2021deep} \cite{wang2021u}, which employ the U-Net architecture developed by Ronneberger et al. \cite{ronneberger2015u}, which has shown to provide superior performance across a wide range of biomedical segmentation tasks \cite{isensee2021nnu}.  Nearly all previous work has shown that U-Net used on rodent models significantly improves the accuracy and reduces the time of rodent brain extract. However, there remain several practical questions to address regarding how such U-net models generalize.  How do they handle multiple image contrasts, data from different scanners, and longitudinal changes due to pathology?

We investigate these questions in the present paper, with the larger goal of developing a brain extraction tool for an image analysis pipeline for the Stroke Preclinical Assessment Network (SPAN) \cite{lyden2022stroke}. SPAN is a multi-center study funded by the National Institutes of Neurological Disorders and Stroke (NINDS) to investigate putative stroke treatments and to address critical issues of rigor, transparency, and reproducibility in preclinical stroke research.  The network includes six research universities and a coordinating center (CC) who manage enrollment of animals, experimental stroke, and blinded and randomized treatment with several candidate cerebroprotectants.  SPAN is an ideal environment for testing the generalizability of preclinical image analytics, as it provides a diverse dataset varied by site, time, and imaging contrast.  SPAN also provides a uniquely large dataset for this evaluation, as our present analysis includes data from 1368 imaging sessions with a total of 5472 image volumes.  In comparison, the largest preclinical brain extraction evaluation was performed by De Feo et al. with 1782 image volumes, collected cross-sectionally from a single imaging center with minimal gross pathology \cite{de2021automated}.  Our experiments examine the performance of a typical U-net CNN model across the wide range of data found in SPAN, including multiple quantitative imaging contrasts, such as T2 and apparent diffusivity coefficient (ADC) maps; multiple time points after experimental stroke; multiple imaging centers with different hardware configurations.  We identify relevant variables that influence the performance of U-net brain extraction models and demonstrate that they can provide excellent performance across these various data conditions.


\section{Methods}
\begin{figure}
\includegraphics[width=\textwidth]{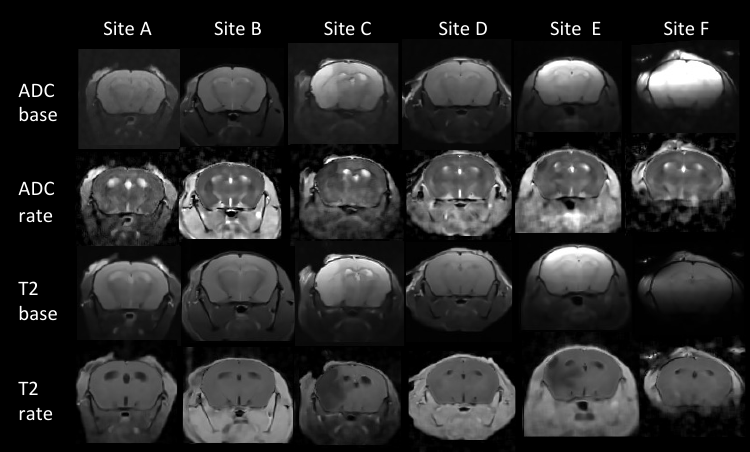}
\caption{Example data showing imaging data from SPAN, which included four quantitative parameter maps collected at six distinct imaging centers. }
\label{fig:images}
\end{figure}

This section focuses on the data used in our experiments, how we constructed our brain extraction models, and our experimental design.  As our work is based around SPAN and ultimately creating a robust image analysis pipeline for the network, our primary goal of this paper is to understand the capabilities and limitations of our brain extraction approach.  We initially developed a traditional rule-based approach, and while this worked in some cases, it was not sufficiently robust for the high-throughput design of SPAN; however, we explored more state-of-the-art U-net approaches.  We describe the data collection scheme of SPAN, the design of these two brain extraction approaches, and how we used SPAN data to comprehensively evaluate the performance of the U-net approach across data types.

{\bf Imaging protocol and data collection:} Data were collected from a mouse model with experimental middle cerebral artery occlusion (MCAO) at day 2 and day 30 after injury. With respective ethics approval, imaging was performed across six imaging centers on Bruker scanners with variable field strengths including 7T, 9.4T, 11.7T, with one site using a surface coil and all others using a volume coil. The multi-parameter imaging protocol included multi-echo T2, and diffusion-weighted MRI (DWI), which were collected at 150 $\mu$m$^2$ coronal in-plane resolution and 500 $\mu$m slice thickness. All sites used three b-values for DWI (0, 500, 1000 s/mm$^2$) and the T2 protocol used either three echoes (0, 45, 75 ms) or ten echoes (equally spaced from 0 to 100 ms). 100 mice were scanned in an initial pilot phase of SPAN to establish SOPs. Following this, SPAN Stage One proceeded to acquire MRI data from 780 animals with a total of 1,368 scanning session, accounting for mortality after injury. All data were routinely uploaded by each site in the DICOM format to the LONI Image Database Archive \cite{Crawford2016-yx} for long term storage and analytics. Example imaging data is shown in Fig. \ref{fig:images}.

{\bf Pre-processing and quality assessment:} While the sequences are similar across sites, there are subtle differences in the image data structure that we first reconcile by parsing DICOM tags, sorting by imaging parameters, fixing image coordinates, converting using dcm2nii, and finally producing a set of matching NIfTI files for each case. We applied adaptive non-local means denoising \cite{manjon2010adaptive} with voxelwise noise estimation, and to account for differences in image grids between scans, we also uniformly resample the images at 150 $\mu$m isotropic resolution using tricubic interpolation. We then perform image quality assessment for each modality by first segmenting foreground and background using Otsu thresholding and computing the signal-to-noise ratio, contrast-to-noise ratio, and signal variance-to-noise variance ratio. We then performed relaxometry to derive quantitative parameter maps, which included a signal baseline and rate of decay for the multi-echo T2 scan (T2$_{base}$ and T2$_{rate}$) and DWI (ADC$_{base}$ and ADC$_{rate}$) scans. For simplicity of presentation, we encoded all T2$_{rate}$ values as the inverse relaxation rate (R2).  These steps were implemented using the Quantitative Imaging Toolkit (QIT) \cite{cabeen2018quantitative}.

{\bf Traditional Rule-based Segmentation Approach:} Our initial brain extraction stage of the SPAN pipeline used a traditional rule-based approach, in which a series of image processing steps to derive a brain mask.  This included the following steps: (1) foreground extraction with an Otsu threshold, (2) histogram-based contrast normalization, (3) edge-preserving smoothing with a non-local means filter \cite{manjon2010adaptive}, (4) gradient magnitude estimation with a Sobel filter bank, (5) thresholding and graph-based segmentation \cite{felzenszwalb2004efficient}, (6) mathematical morphology including opening, closing, and filling, and (7) regularization with a Markov random field \cite{felzenszwalb2006efficient}.  We applied this procedure to the ADC$_{base}$ contrast because it showed the least lesion contrast. This approach was effective in a large proportion of cases (over 80\%), and most failures were due to partial volume effects that blurred the skull boundary, resulting in either small errors on the superficial surface of the brain or catastrophic errors in which the brain mask ``leaked'' to the rest of the head.  We also experiment with the AFNI rodent segmentation method 3dSkullStrip \cite{cox1996afni} as well as BET \cite{smith2000bet} with the input parameters ({\it -shrink\_fac .8 -rat}) and ({\it -f 0.8 -R -m}) respectively; however after quality control, we did not find these to provide improved performance over the described rule-based approach. This challenges pose a limitation for its use in SPAN, as we needed to preserve as much data as possible, so we used this approach to bootstrap a more accurate and modern approach using U-nets.


{\bf Modern Neural Network Approach:} We developed a brain extraction tool using a CNN neural network approach with a U-net architecture proposed by Ronneberger et al. \cite{ronneberger2015u}, which has demonstrated superior performance in previous rodent imaging studies \cite{Hsu2020-ku} \cite{De_Feo2021-xy}.  Our model was implemented in PyTorch and included five contraction/expansion stages, a kernel size of 64, and an input resolution of 128x128.  We trained a single 2D U-net with data from all image planes, and during inference, we applied the 2D U-net three times to include every possible image plane, and we finally took the average prediction from the three applications to each voxel to obtain the final prediction.  We trained our model with the ADAM optimizer with a cross-entropy loss, a learning rate of 0.0001, a batch size of 20, and included albumentations-based data augmentation for translation, rotation, scaling, contrast, and deformations \cite{buslaev2020albumentations}. Our system was a Lambda Labs workstation with Ubuntu 20.04, a AMD Ryzen Threadripper 3970X 32-Core CPU, and an Nvidia 1080 Ti 12GB GPU.  We created our training dataset from the best performing results from the our previous rule-based approach.  In particular, we selected training examples from 180 cases and hold-out testing and validation examples from 30 cases; each was split roughly evenly between imaging centers and time points.  We trained the model for 10 epochs on the training data, selected the best model with the validation data, and finally evaluated hold-out performance with the test data.  We report our accuracy estimates with the Dice coefficient \cite{dice1945measures}.  This described model is the focus of our evaluation experiments, which are described in the next section. 


{\bf Experimental design:} Because CNNs are black-box approaches by nature, a major question is how they perform across different conditions. Substantial previous work has tested how different types of U-net architecture affect segmentation performance \cite{de2021automated} \cite{isensee2021nnu}.  In contrast, we sought to understand how U-net performance varies across different types of data. SPAN provides a wide range of data types, varying across imaging contrasts, scanner hardware, and longitudinal time points, and we investigated how a typical U-net architecture performs relative to these different conditions.  This is important both for SPAN, but it also may provide useful knowledge about how these models generalize to other preclinical studies and when their scope is expanded.  Our experiments systematically tested these factors using the SPAN data as follows.  We fit eight U-net models on different combinations of data and investigated how different data characteristics relate to segmentation performance.  Among these eight models, we trained models that include all four quantitative parameters as multi-channel input (R2$_{base}$, R2$_{rate}$, ADC$_{base}$, and ADC$_{rate}$) as well as a single-channel model trained on the concatenated whole of these image contrasts.   We also examined performance of four models trained separately on each image contrast.  Finally, we investigated the ability of the model to generalize to new sites by holding out data from three of the sites.  Specifically, for this site-based test, the training and validation data were from three sites (B, D, F) and the testing data was only from three other distinct sites (A, C, E), thus it is designed to indicate how well the model might generalize to unseen data from sites added to SPAN in the future.  We conducted this site-based test two, for both multi- and single-channel models.  Note: because there are four image contrasts, the single-channel models has four times the data as the multi-channel models.  As a final evaluation step, we applied our brain extraction tool to the entirety of the SPAN Stage One dataset (N = 1368) and qualitatively assessed the results to determine if they pass quality control; we report this outcome as the ``failure rate''. The 8 models we trained will be referred to as such: M-full-multi (MFM), M-full-single (MFS), M-half-multi (MHM), M-half-single (MHS), M-contrast-T2-base (MT2B), M-contrast-T2-rate (MT2R), M-contrast-ADC-base (MADCB), and M-contrast-ADC-rate (MADCR).

\section{Results and Discussion}

In this section, we present results from experimental evaluation of the U-net model across data types.  We first examine results from a multi-channel U-Net model trained on the full dataset, and then compare it to a single channel model where each scan is fed individually into the model.  We then report how these same models perform for specific sites and specific time points.  We then report how restricted models perform, i.e. those looking a model restricted to train on only half of the sites (with the remaining half left for testing), as well as models trained on specific imaging contrasts.

\begin{table*}
\centering
\label{table:results}
\ra{1.3}
\caption{Quantitative results from our experiments.  The top table shows results from two models: one trained using multi-channel input with four contrasts and the other trained with a single channel input taking each contrast separately.  The first column shows the overall results and the following columns show results broken down by site and time point.  The bottom table shows results for more specific models which were trained on a subset.  The first four columns show results from training the model on individual image contrasts. The last two columns show results from training a model on data from sites B, D and F, and then testing it on sites A, C, and E to gauge generalization across sites.}

\vspace*{1em}
\begin{tabular}{@{}lcccccccccccc@{}}\toprule
& & \phantom{abc}& \multicolumn{6}{c}{Split by Imaging Site} & \phantom{abc} & \multicolumn{2}{c}{Split by Time}\\
 \cmidrule{4-9} \cmidrule{11-12}
Channels & All & & A & B & C & D & E & F && Day 2 & Day 30 \\ 
\midrule
Multi & 0.964 && 0.974 & 0.976 & 0.980 & 0.972 & 0.965 & 0.914 && 0.953 & 0.974 \\
Single & 0.957 && 0.967 & 0.967 & 0.973 & 0.952 & 0.964 & 0.919 && 0.946 & 0.968 \\
\bottomrule
\end{tabular}
\newline
\vspace*{1em}
\newline
\begin{tabular}{@{}lccccccc@{}}\toprule
& \multicolumn{4}{c}{Models for Each Contrast} & \phantom{abc}& \multicolumn{2}{c}{Half-site Restricted Model} \\
\cmidrule{2-5} \cmidrule{7-8} 
& ADC$_{base}$ & ADC$_{rate}$ & R2$_{base}$ & R2$_{rate}$ && Multi-channel & Single-channel \\
\midrule
Dice Score & 0.964 & 0.958 & 0.935 & 0.954  && 0.953 & 0.949 \\
\bottomrule
\end{tabular}
\end{table*}

\begin{figure}[!t]
\includegraphics[width=\textwidth]{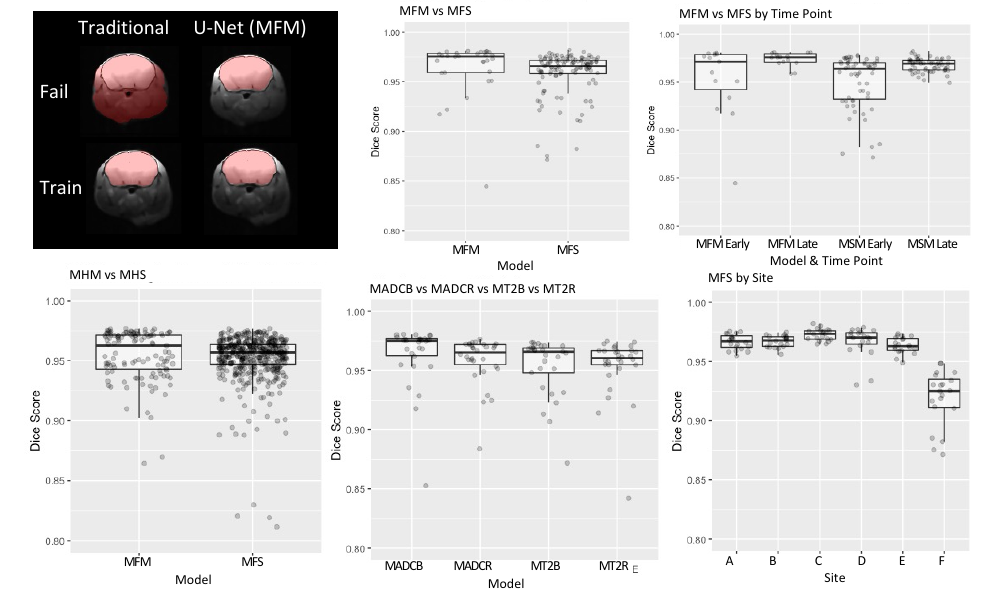}
\caption{Results (left to right): 1) Comparison of traditional segmentation and MFM U-Net on a fail and a training case; 2) MFM vs MFS total dice scores; 3) MFM vs MFS broken down by time point (early vs late); 4) MFM and MFS total dice scores; 5) All 4 M-Single Contrast models (MADCB,MADCR,MT2B, MT2R); 6) MFS broken down by site.}
\label{fig:plots}
\end{figure}


Our results are summarized quantitatively in Table 1 and visualized in Fig. \ref{fig:plots}. M-full-multi had the highest performance with a Dice score of 0.964. The M-full-single had only slightly lower performance than the multi-channel model with a Dice of 0.957. Looking at the data qualitatively, both M-full-multi and M-full-single model had a 0\% failure rate. This trend was consistent across sites and time points. When broken down by site, the Dice scores reflected image quality, with Site F having the highest SNR and the lowest dice scores of 0.914 and 0.919 for the multi and single channel models respectively. This site was the sole site with a surface receive coil (the other sites had volume coils), which may also explain the lower performance. Considering the restricted models, training the models on half (B,D,F) of the sites and testing it on the other half (A,C,E) resulted in a mean Dice scores of 0.953 (M-half-multi) and .954 (M-half-single), indicating that both models can perform robustly on data from sites added to SPAN in the future, without being trained on data from those sites. Considering the restricted models for individual contrasts, M-contrast-ADC-base, M-contrast-ADC-rate, M-contrast-T2-base, M-contrast-T2-rate (Table 1) also had robust Dice scores, with T2 baseline being the lowest at 0.935, indicating that the model can be trained on a single contrast and still perform robustly on that contrast.  In both M-full-multi and M-full-single, there was an increase in performance between the early and late time points. This is likely because the brain has a more regular appearance at the late time point 30 days post injury and consists of more healthy tissue, while the early time point 2 days post stroke, it has greater injury and morphometric abnormality, leading to slightly lower performance.

We selected a group of 206 cases that failed catastrophically when processed by the rule-based brain extraction approach, and we looked at the qualitative performance of the U-net in these specific cases. The M-full-multi successfully extracted all of these cases, and two independent quality checks of the segmented data from the first stage of the SPAN study (1368 4-channel scans) showed a 4\% failure rate on large multi-site preclinical mouse ADC and T2 datasets with stroke pathology.  Some of the cases that failed quality control were excluded due to motion artifact, operator error, etc.; because it includes these other factors, it is more general than only the failures from the brain extraction step.

Considering our findings in relation to previous work, the Dice scores of our models are comparable to an existing study of the U-Net’s performance on large preclinical data conducted by De Feo et al. in 2021 while evaluating network architectures. The model of De Feo et al. achieved a brain mask Dice score of 0.978 when tested on 1782 T2 volumes of healthy and Huntington mice  \cite{de2021automated}. It should be noted that while the Huntington mice are a disease model, they lack the morphometric abnormality found in stroke cases, so we could reasonable expect the performance to be higher.  Our results show that when the U-Net is evaluated on thicker slices and a variety of contrasts, image channels, time points, and sites, it maintains the robustness demonstrated by De Feo et al.

\noindent {\bf Conclusions:}  We evaluated the U-Net neural network model’s ability to segment a large dataset from a multi-site preclinical rodent stroke imaging study, exploring a myriad of factors related to data quality and character. We have demonstrated that the U-Net performs reliable brain extraction on mice data collected from various imaging hardware, multiple time points, with varying contrasts and field strengths where traditional methods failed. Quality control of 1,368 scans segmented by our model has shown that this technique can be successfully used to create a robust pipeline for mice brain extraction in high throughput preclinical imaging studies. We also found that a single-channel model of the U-Net is nearly as robust as the multi-channel version, allowing flexibility in reducing the scanning protocol in subsequent stages SPAN going forward. Future opportunities include testing the U-Net’s generalization to rats, and specimens that are obese, aged, and female.  We may further evaluate the speed of this approach on various hardware available to labs, i.e. comparing performance on single GPU machine or CPU-based grid computing environments. Looking forward, our model can help inform the design of preclinical studies and potentially improve their scalability and reliability in the future. We provide an open-source implementation and trained models of our method online at \url{http://github.com/cabeen/neu-net}.\\ 

\noindent {\bf Acknowledgements:} Work reported here was supported by grant NS U24 NS113452 (PL). RPC is supported by the CZI Imaging Scientist Award Program, under grant number 2020-225670 from the Chan Zuckerberg Initiative DAF, an advised fund of Silicon Valley Community Foundation.



%




%
%
%
%

\bibliographystyle{splncs04}
\bibliography{paper}

%
%
%
%
%
%
\end{document}